# Trie Constraints and Hierarchy-Aware Semantic Alignment for HS Code Prediction with Small Language Models

Minseop Kim[a], Taekhyun Park[b], Kikun Park[c], Hyerim Bae[a*]

[a]Major in Industrial Data Science & Engineering, Department of Industrial Engineering, Pusan National University, Republic of Korea
[b]Graduate School of Data Science, Pusan National University, Republic of Korea
[c]Port Solution Group, Digital Team, CyberLogitec, Republic of Korea

**Abstract**

Harmonized System (HS) code prediction (HSP) from commodity text is essential to international trade, and its importance continues to grow in port logistics. Recently, large language models (LLMs) have been actively investigated for this task, owing especially to their strong language-understanding capabilities. However, their high computational cost limits deployment in constrained environments such as container terminals. Small language models (SLMs) offer a practical alternative, but their smaller scale makes them prone to generating invalid HS codes and to overlooking the hierarchical semantics between commodity text and HS codes. To address these limitations, this study proposes TRIE-HSA, which combines trie-constrained token prediction with hierarchy-aware semantic alignment (HSA). This framework constrains the SLM to predict only valid digits under the HS taxonomy and aligns commodity text representations with the hierarchical structure of HS codes. In extensive experiments on data collected from an operational container terminal, TRIE-HSA improved average HS6 accuracy by 49.96% over zero-shot inference and exceeded the strongest task-specific benchmark by 11.94%. These results demonstrate that accurate and structurally valid HSP is achievable with fewer than 10 billion parameters. Therefore, TRIE-HSA offers a practical basis for deployment of HSP in port logistics operations that cannot support large-scale LLMs.

## 1. Introduction

The Harmonized System (HS) is an internationally standardized commodity classification system that serves as the basis for tariff determination, regulatory compliance, and customs clearance worldwide [1,2]. Accurate HS code classification is essential to international trade, because misclassification can lead directly to operational losses resulting from inclusion of excessive duties, regulatory violations, and logistical inefficiencies [3,4]. In this light, HS code prediction (HSP) has been an active area of research across various fields [4,5]. However, HSP is inherently difficult, for two reasons. First, the HS taxonomy is highly granular and hierarchical. The first two digits identify one of 96 chapters (HS2), the four digits identify one of 1,228 headings (HS4), and the complete six-digit code identifies one of 5,612 subheadings (HS6) [6]. Second, commodity text is ambiguous. Different commodities are often described in highly similar terms, while the same commodity can be expressed in substantially different ways [4,7]. These difficulties demand considerable domain expertise and leave the classification process susceptible to human error.

More recently, studies have explored the language-understanding capabilities of large language models (LLMs) for HSP [2,8,9]. LLM-based methods can capture semantic information from unstructured commodity text more effectively than conventional approaches, thereby broadening the application of HSP to port logistics. Kim et al. [8] predicted HS codes from commodity text and used the resulting predictions to improve container terminal productivity. Their findings not only demonstrate that HSP can contribute directly to real-world port operations but also highlight the potential for further research on LLM-based HSP in port logistics. However, LLMs require substantial computational resources, whereas many port logistics facilities operate in resource-constrained environments. Therefore, practical deployment requires HSP methods based on small language models (SLMs) offering performance comparable to that of LLMs [2,8].

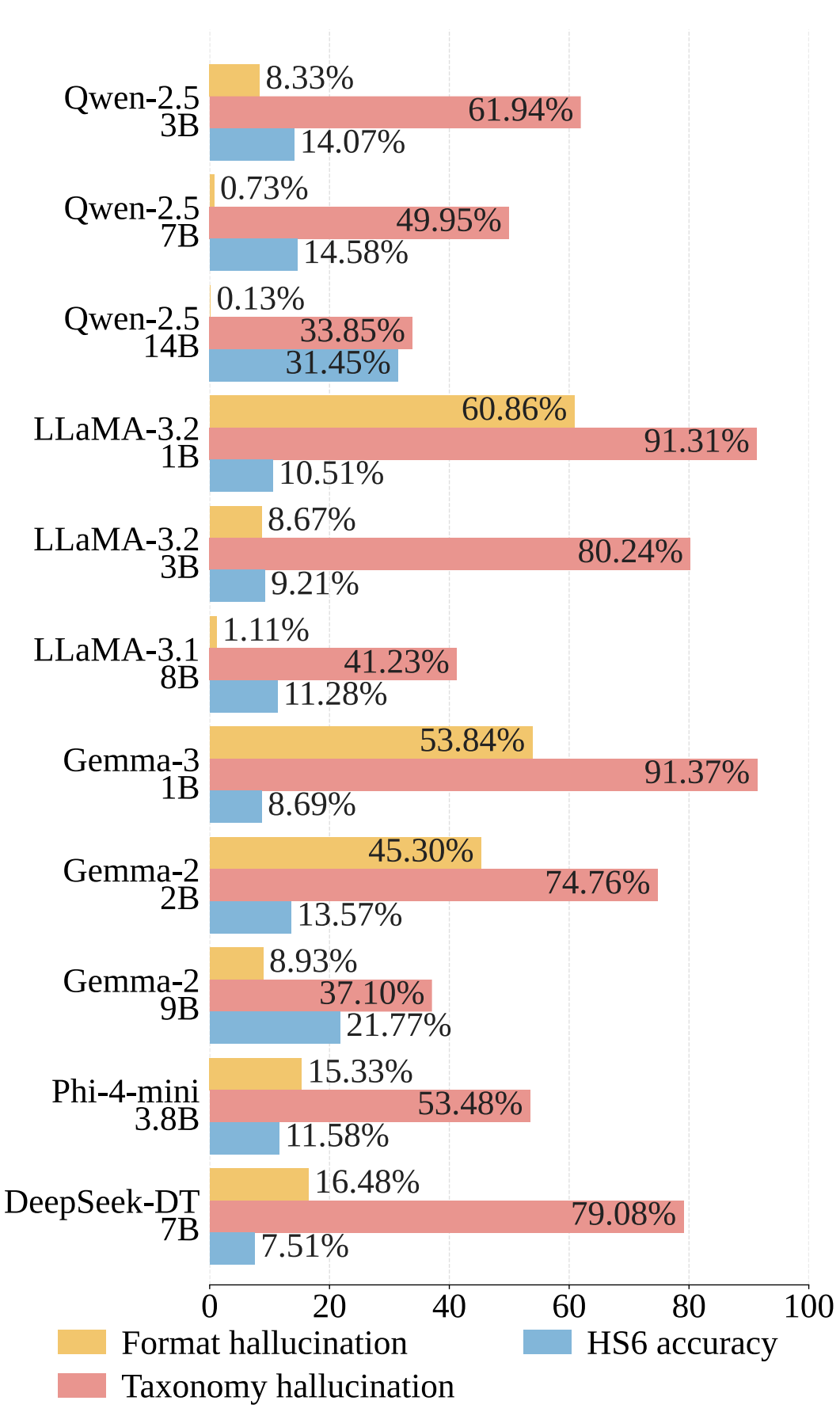

**Fig. 1. HSP performance and hallucination across SLMs**

As illustrated in Fig. 1, applying an existing SLM directly to HSP without task-specific fine-tuning has clear limitations. Here, format hallucination denotes an output that is not a six-digit code, and taxonomy hallucination denotes a six-digit code absent from the HS taxonomy. Even the best-performing SLM reaches only 31.45% HS6 accuracy and generates frequent hallucinated codes, indicating that a pretrained SLM alone is insufficient to guarantee taxonomically valid outputs and to achieve high performance. These limitations motivate two requirements for effective SLM-based HSP, constraining predictions to valid HS codes and aligning commodity text with the hierarchical semantics of the HS taxonomy.

To meet these requirements, this study proposes TRIE-HSA, a framework that integrates trie-constrained token prediction with hierarchy-aware semantic alignment (HSA). The proposed framework adopts a trie mechanism that shares common prefixes among HS codes in order to constrain each prediction step of the SLM to valid digits. HSA, meanwhile, employs learnable prototypes at the chapter and heading levels to align commodity text representations with the HS code hierarchy. These two components enable TRIE-HSA to eliminate hallucinations while incorporating hierarchical semantic information into SLM representations. This framework was evaluated using commodity text collected from a real-world container terminal. The results demonstrate that TRIE-HSA outperforms existing SLMs and HSP methods.

The remainder of this paper is organized as follows. Section 2 reviews the relevant previous studies on HSP. Section 3 describes the proposed TRIE-HSA methodology. Section 4 presents the experimental design and results. Section 5 discusses the practical implications, limitations, and future research directions. Finally, Section 6 concludes the paper.

## 2. Related Work

### *2.1 HS Code Prediction (HSP)*

Early research on HSP relied primarily on text mining and machine learning. Ding et al. [10] proposed a Background Net-based approach that processes commodity text by learning word co-occurrence frequencies. Altaheri and Shaalan [6] represented user-provided product descriptions using term frequency-inverse document frequency (TF-IDF) and compared several machine-learning algorithms for HSP, reporting that a linear support vector machine (LSVM) achieved the best performance. Spichakova and Haav [11] highlighted the short and noisy nature of real-world commodity text and proposed a combined measure that pairs textual cosine similarity with taxonomy-based similarity between HS codes. These early machine-learning approaches relied on manually engineered features based on keywords or statistical properties, and therefore had difficulty capturing the semantic context of unstructured text.

With advances in deep learning, HSP research has developed in two broad directions. The first improves the semantic understanding of commodity text through network architectures, text-embedding models, and contrastive learning. Lee et al. [12] improved HSP performance by fine-tuning KoELECTRA and combining it with sentence retrieval, while Amel et al. [13] investigated a multimodal approach that jointly uses textual and visual information. He et al. [14] combined a convolutional neural network (CNN) and BERT to predict HS codes absent from the training data, which conventional text-similarity systems cannot utilize. Their framework outperformed traditional machine-learning models, though its performance depended heavily on the quality of the source data. Zhou et al. [7] proposed a shallow-structured convolutional neural network (SSCNN) that learns the core meaning of ambiguous commodity text. However, this approach assumes that declaration data is structured across multiple fields and generalizes poorly when only a single commodity-text field is available. Sun et al. [15] and Qi et al. [4] represented attributes in commodity text as graphs and demonstrated that knowledge graphs improve HSP performance. Anggoro et al. [16] additionally combined supervised contrastive learning with Sentence-BERT to increase semantic similarity among commodity texts assigned to the same HS code. Collectively, these studies advanced the semantic representation of commodity text from single-model encoders toward multimodal, graph-based, and contrastive approaches.

The second research direction exploits the semantic information that an HS code carries at the chapter, heading, and subheading levels, and predicts these levels hierarchically. This research was a response to the dilemma that predicting the six digits of an HS code as a single class requires selecting one label from several thousand candidates while ignoring the underlying hierarchy [7]. Bihn et al. [17] combined an attention mechanism with an RNN and introduced four softmax layers to reflect the HS code hierarchy in prediction. Chen et al. [18] proposed a framework that translates commodity text into HS codes using an RNN- and attention-based neural machine translation (NMT) mechanism. Shubham et al. [19] proposed a hybrid framework that predicts the chapter level (HS2) with BERT and then predicts the six-digit HS code by cosine similarity. This framework improved accuracy by approximately 16% over predicting the six digits as a single class. These findings suggest that explicitly modeling the hierarchy between commodity text and HS codes can improve HSP performance.

The recent emergence of LLMs with powerful language-understanding and generation capabilities has opened new possibilities for HSP. LLMs improve HSP performance because they represent the meaning of unstructured commodity text more effectively than the models used in previous studies [3,4,20,21]. Accordingly, LLM-based HSP research has increased. Marra De Artiñano et al. [22] conducted the first systematic evaluation of GPT-3.5 for HSP and achieved accuracy ranging from 60 to 90%, depending on the HS code level. In contrast to previous studies, this evaluation suggests that LLMs offer a promising approach to HSP across multiple datasets without task-specific preprocessing or training data. Yuvraj and Devarakonda [2] proposed ATLAS, a supervised fine-tuned LLaMA-3.3-70B, and reported performance exceeding that of GPT-5-Thinking. Lan et al. [9] applied recent LLMs and an agent-based framework to HSP, demonstrating the potential of LLMs to actively retrieve and reason over extensive external knowledge. These studies indicate that LLMs may establish a new paradigm for HSP.

Recent investigations into port logistics have increasingly used HSP to improve operational productivity. Xie et al. [23] proposed a method that predicts standard international trade classification (SITC) codes from the commodity text of

import containers and used these predictions to train a machine learning (ML) model that classifies their out-terminals. The study demonstrated that prediction of SITC codes improved ML model performance and reduced costs by decreasing unnecessary equipment movements in port operations. Furthermore, Kim et al. [8] adopted Gemini 2.5 to predict HS codes from the commodity text of import containers and applied the predictions to container dwell-time prediction tasks. The study reported that the predicted HS code was the most important variable for dwell-time prediction and demonstrated that incorporating these predictions into container-stacking strategies reduced yard-crane re-handling relative to existing operations. This work exemplifies how LLM-based HSP can support core decisions in port operations. Nevertheless, the high cost of LLMs remains a major limitation in resource-constrained port environments [8]. Therefore, adapting SLMs for deployment in port operations is an important research direction. To this end, this study proposes TRIE-HSA, which combines trie-constrained token prediction with hierarchy-aware semantic alignment.

### *2.2 Constrained Token Prediction Mechanism*

Many natural language processing (NLP) tasks require a model to produce output that follows a predefined structure. Research on constrained generation began before the emergence of LLMs, in the field of sequence generation. Hokamp and Liu [24] placed lexical constraints on beam search so that required words appear in the output. Satisfying such constraints becomes more difficult with LLMs, because they are trained to generate text and select from tens of thousands of tokens at each generation step [25–27]. Lu et al. [26] observed that even supervised LLMs fail to follow basic prediction constraints, and proposed lexical constraints that force specific words to be included in or excluded from an output. Scholak et al. [27], Geng et al. [25], and Willard and Louf [28] guaranteed valid output formats by removing tokens that violate a predefined structure from the probability distribution at every generation step. Constrained token prediction has also been applied to hierarchical code systems. Coutinho et al. [29] trained a relatively small LLaMA model on death certificate text and applied constrained decoding at inference so that only valid ICD-10 codes are generated. This generative approach reached accuracy comparable to that of a BERT encoder classifier at the chapter, block, and full-code levels of the ICD hierarchy.

Several studies have used a trie mechanism to determine the tokens available in tasks that allow only a constrained output, such as entity retrieval and generative recommendation. A trie groups the valid items by their shared prefixes, so that the generated prefix determines the candidates for the next output. De Cao et al. [30] fine-tuned a pretrained sequence-to-sequence model to generate entity names autoregressively. This study adopted a trie over the entity names in a knowledge base and constrained each step to the continuations of the generated prefix, so that the model returned an existing entity. Su et al. [31] noted that the decoding of an LLM cannot natively constrain its output to a predefined set of valid items, and applied a trie mechanism at each decoding step. Combining a trie with LLM decoding guaranteed valid outputs without modifying the model and introduced negligible computational overhead at inference. These advantages carry over to SLM-based HSP, since each digit of an HS code from chapter to subheading follows from the preceding prefix. Therefore, this study proposes trie-constrained token prediction so that an SLM predicts only valid HS codes at little additional cost.

### *2.3 Prototype-Contrastive Learning*

Prototype-contrastive learning defines the semantic center of each label as a prototype and performs contrastive learning against these prototypes. This design avoids the large number of positive and negative pairs that conventional contrastive learning requires, and so it is used mainly when the training data are imbalanced or scarce. Song et al. [32] introduced category prototypes into a supervised contrastive loss so that every sample obtains positive and negative signals under class imbalance, and their method remained stable at small batch sizes. Cui et al. [33] learned a prototype for each class from a few training instances. The prototypes replaced hand-designed label words and served as the criterion for few-shot classification. Wang et al. [34] extended prototypes to multi-label settings wherein a sample belongs to several categories and aligned label-specific text representations with the prototype of each label. These studies show that prototypes provide stable training signals from limited and imbalanced data. However, these methods rarely capture the hierarchical relationships among categories.

Other studies have sought to incorporate label hierarchies into representation learning. Wang et al. [35] observed that conventional methods separately encode and combine text and label hierarchies, resulting in the same hierarchical representation regardless of the input. They directly injected hierarchical information into a text encoder through contrastive learning with positive samples constructed under the guidance of the label hierarchy. Yu et al. [36] extracted label-specific embeddings using multi-head attention and combined instance-level and label-level supervised contrastive learning. Their method aligned representations of similar samples within the hierarchy more closely and incorporated semantic relationships among labels into the representation space, thereby improving hierarchical multi-label classification. Without requiring explicit hierarchical labels, Kim et al. [37] proposed a regularization method that approximates the latent semantic hierarchy of data using hierarchical proxies learned in hyperbolic space. Zhang et al. [38] proposed a hierarchy-aware and label-balanced model to address both insufficient hierarchy awareness and label imbalance in hierarchical text classification.

To apply an SLM to HSP, two conditions need to be satisfied. First, prototypes at the same level need to be separated semantically from one another (within-level). Second, each heading prototype needs to stay close to its parent chapter prototype (cross-level). Meeting these conditions is difficult in HSP, because the prediction space contains 5,612 subheadings while the training data are imbalanced, so many category pairs have too few examples to learn from. Prototype-based learning suits this setting, since it contrasts each sample against one semantic center per category rather than against sample pairs. Accordingly, this study applies hierarchy-aware semantic alignment to an SLM. Learnable prototypes serve as the semantic centers of chapters and headings. They are separated from one another at each level, and each heading prototype is aligned with its parent chapter prototype.

## 3. Methodology

This section defines the SLM-based HSP problem for the current study and describes the proposed TRIE-HSA framework. Fig. 2 presents an overview of the framework, and Table 1 summarizes the notation.

### *3.1 Problem Formulation: Autoregressive HSP*

Given a commodity text, a tokenizer segments it into the token sequence $x$ (Eq. (1)), where each token $x_t$ is a word or subword that a language model processes as a minimal unit and $T$ denotes the token length.

$$x = (x_1, \dots, x_T) \tag{1}$$

From this sequence, an SLM $f_\theta$ autoregressively predicts the next token, which is the standard way in which LLMs generate sentences and is referred to as next-token prediction (NTP) [39]. Because the output of SLM-based HSP is a six-digit HS code, this study reformulated the task as autoregressively predicting each digit as a token. Therefore, an HS code is denoted by a six-digit sequence (Eq. (2)), where each digit $y_p$ takes a value in $(0, \dots, 9)$. The code carries hierarchical meaning in two-digit intervals, where the first two, four, and six digits correspond to the chapter, heading, and subheading levels, respectively.

$$y = (y_1, y_2, \dots, y_6) \tag{2}$$

The SLM predicts the six digits autoregressively, where $\hat{y}_p$ denotes the token predicted at position $p$. At each position $p$, the input sequence $\mathcal{S}_p$ comprises the token sequence $x$ and the previously predicted tokens (Eq. (3)), constructed recursively from $\mathcal{S}_1 = x$ by appending each predicted token. Hence, $\mathcal{S}_p$ contains $T + p - 1$ tokens.

$$\mathcal{S}_p = (x_1, \dots, x_T, \hat{y}_1, \dots, \hat{y}_{p-1}) \tag{3}$$

The backbone $f_\theta$ encodes $\mathcal{S}_p$ and yields the hidden state $h_p$ (Eq. (4)). Through self-attention mechanism, $h_p$ aggregates the semantic information of the entire sequence.

$$h_p = f_\theta(\mathcal{S}_p) \in \mathbb{R}^H \tag{4}$$

The output projection head $W_{head}$ maps $h_p$ onto the logit vector $logit_p$ over the vocabulary $V$, which is the set of all possible output tokens (Eq. (5)).

$$logit_p = W_{head} h_p \in \mathbb{R}^V \tag{5}$$

Applying softmax normalizes $logit_p$ into the probability distribution $q_p$ (Eq. (6)). Finally, the token with the highest probability in $q_p$ is selected as $\hat{y}_p$ (Eq. (7)), and this procedure repeats until the sixth digit.

$$q_p = softmax(logit_p) \in \mathbb{R}^V \tag{6}$$

$$\hat{y}_p = \arg\max_v q_{p,v} \tag{7}$$

Conventional NTP encounters a structural limitation for SLM-based HSP. Because the output space of an SLM is determined by the vocabulary $V$, the model can predict non-numeric tokens invalid under the HS taxonomy. Therefore, this NTP process cannot guarantee a valid HS code.

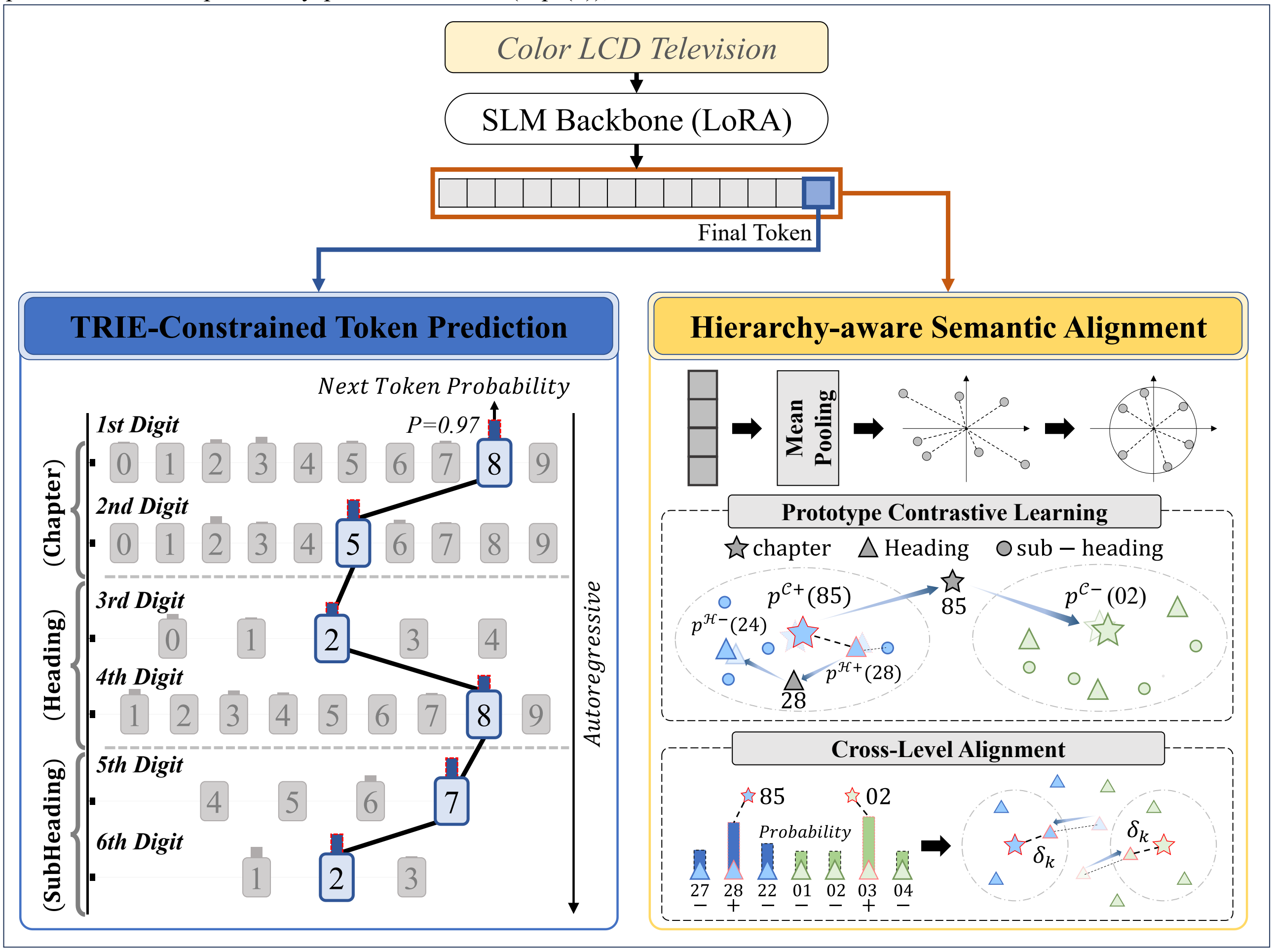

**Fig. 2. Overview of TRIE-HSA framework**

**Table 1. Notation**

| Symbol | Definition | Symbol | Definition |
|---|---|---|---|
| $p$ | Predicted digit position index $p \in 1, \dots 6$ | $T$ | Token length |
| $i$ | Chapter index $i \in \mathcal{C}$ | $V$ | Vocabulary, the set of possible output tokens |
| $j$ | Heading index $j \in \mathcal{H}$ | $\mathcal{C}$ | Set of chapters |
| $\pi(j)$ | Parent chapter of heading $j$ | $\mathcal{H}$ | Set of headings |
| $v$ | Vocabulary index | $\mathcal{X}_i^{\mathcal{C}}$ | Subset of commodity texts of chapter $i$ |
| $x$ | Tokenized commodity text | $\mathcal{X}_j^{\mathcal{H}}$ | Subset of commodity texts of heading $j$ |
| $d$ | Single digit of HS code | $\mathcal{T}$ | HS trie constructed from HS taxonomy |
| $q$ | Probability distribution | $\mathcal{D}_p$ | Valid digit set at position $p$ |
| $y$ | Actual HS code | $\mathcal{S}_p$ | Input sequence at position $p$ |
| $\hat{y}$ | Predicted HS code | $\mathcal{M}_p$ | Trie mask vector at position $p$ |
| $f_\theta$ | Backbone SLM | $P_i^{\mathcal{C}}$ | Chapter prototype |
| $h_p$ | Hidden state at position $p$ | $P_j^{\mathcal{H}}$ | Heading prototype |
| $h_t$ | Hidden state of token $x_t$ | $\delta_j$ | Heading offset |
| $\bar{h}_x$ | Mean-pooled representation from $f_\theta(x)$ | $\mathcal{N}$ | L2-normalization |

*3.2 Trie-constrained Token Prediction*

To address the limitation discussed in Section 3.1, we introduce a trie mechanism. A trie is a tree-based data structure mainly used to store and search large collections of strings based on their common prefixes. Because the generated prefix determines the candidate digits at the next position, the trie mechanism can constrain the SLM to predict only valid HS codes.

To this end, we construct an HS trie $\mathcal{T}$ from the HS taxonomy. Given the previously predicted prefix $\hat{y}_{<p} = (\hat{y}_1, \dots, \hat{y}_{p-1})$, $\mathcal{T}$ returns the valid digit set $\mathcal{D}_p$ at position $p$ (Eq. (8)). At the first position, the prefix is empty and $\mathcal{T}$ returns all ten digits.

$$\mathcal{D}_p = \mathcal{T}(\hat{y}_{<p}) \tag{8}$$

Next, we define the trie mask $\mathcal{M}$, which assigns 0 to tokens in $\mathcal{D}_p$ and $-\infty$ otherwise, to constrain the output space of the SLM (Eq. (9)). Here, $p$ is the position to be predicted and $v$ indexes each token in the vocabulary $V$. Adding $\mathcal{M}_{p,v}$ to the logit vector $logit_p$ yields the masked logit vector $z_p$ at position $p$ (Eq. (10)).

$$\mathcal{M}_{p,v} = \begin{cases} 0, if\ v \in \mathcal{D}_p \\ -\infty, otherwise \end{cases} \tag{9}$$

$$z_p = logit_p + \mathcal{M}_{p,v} \in \mathbb{R}^V \tag{10}$$

This process preserves the logits of the tokens in $\mathcal{D}_p$ and suppresses invalid tokens for HSP, so that the SLM predicts only valid HS codes. This constraint also concentrates the probability mass within $\mathcal{D}_p$, supporting high-performing SLM-based HSP when combined with fine-tuning. We refer to this approach as trie-constrained token prediction.

During fine-tuning, we adopt the cross-entropy loss generally used for training an SLM. This loss encourages the probability distribution of the masked logit vector $z_p$ to match the ground-truth digit $y_p$. Because an HS code consists of six digits, we average the loss over the six positions and define the trie-constrained loss $\mathcal{L}_{trie}$ as follows (Eq. (11)).

$$\mathcal{L}_{trie} = \frac{1}{6}\sum_{p=1}^{6} CE(z_p, y_p) \tag{11}$$

The digit set $\mathcal{D}_p$ is retrieved using the previously predicted prefix $\hat{y}_{<p}$ from $f_\theta$. Because $f_\theta$ predicts each digit autoregressively, $\mathcal{L}_{trie}$ requires correct predictions at all six positions. However, $f_\theta$ cannot guarantee this during fine-tuning, and a prediction error at an earlier position leads subsequent positions to retrieve an incorrect $\mathcal{D}_p$. Therefore, we apply teacher forcing, which retrieves $\mathcal{D}_p$ using the ground-truth prefix $y_{<p}$ rather than the predicted prefix. This process guarantees that $\mathcal{D}_p$ contains the ground-truth digit at every position and stabilizes fine-tuning with the trie-constrained loss $\mathcal{L}_{trie}$.

*3.3 Hierarchy-Aware Semantic Alignment*

In addition to trie-constrained token prediction, SLM-based HSP requires learning the semantic relationship between commodity text and the HS taxonomy [16]. As discussed in Section 2.3, categories at the same level need to be distinguishable in the representation space (within-level), and each heading needs to remain semantically close to its parent chapter (cross-level). To meet these two conditions, we introduce learnable prototypes representing the semantic centers of chapters and headings and propose hierarchy-aware semantic alignment (HSA).

*3.3.1 Prototype Construction*

HSA first constructs initial prototypes from the pretrained SLM $f_\theta$ to exploit its representation capability. We input every commodity text $x$ except the validation and test data into $f_\theta$ and obtain the mean-pooled representations $\bar{h}_x$ by averaging its token-level hidden states $h_t$ over the token length $T$ (Eq. (12)). This representation captures the meaning of a single commodity text.

$$\bar{h}_x = \frac{1}{T}\sum_{t=1}^{T} h_t \tag{12}$$

To initialize the prototypes at each level, we aggregate these representations at the chapter and heading levels. The chapter centroid $\mu_i^{\mathcal{C}}$ and heading centroid $\mu_j^{\mathcal{H}}$ are the L2-normalized means of $\bar{h}_x$ over the commodity texts of chapter $i$ and heading $j$ (Eqs. (13) and (14)), where $\mathcal{X}_i^{\mathcal{C}}$ and $\mathcal{X}_j^{\mathcal{H}}$ denote the corresponding sets of commodity texts.

These centroids represent the meaning of each chapter and heading.

$$\mu_i^{\mathcal{C}} = \mathcal{N}\left(\frac{1}{|\mathcal{X}_i^{\mathcal{C}}|}\sum_{x\in\mathcal{X}_i^{\mathcal{C}}} \bar{h}_x\right) \tag{13}$$

$$\mu_j^{\mathcal{H}} = \mathcal{N}\left(\frac{1}{|\mathcal{X}_j^{\mathcal{H}}|}\sum_{x\in\mathcal{X}_j^{\mathcal{H}}} \bar{h}_x\right) \tag{14}$$

From these centroids, we construct the prototypes of the two levels. First, the chapter prototype $P_i^{\mathcal{C}}$ is initialized with its chapter centroid (Eq. (15)).

$$P_i^{\mathcal{C}(0)} = \mu_i^{\mathcal{C}} \tag{15}$$

A heading prototype needs to carry semantics similar to those of its parent chapter prototype and to preserve this hierarchy during fine-tuning. Therefore, we introduce a heading offset $\delta_j$, initialized as the difference between the heading and chapter centroids (Eq. (16)).

$$\delta_j^{(0)} = \mu_j^{\mathcal{H}} - \mu_i^{\mathcal{C}} \tag{16}$$

The heading prototype $P_j^{\mathcal{H}}$ is then parameterized by adding this offset to its parent chapter prototype $P_i^{\mathcal{C}}$ (Eq. (17)).

$$P_j^{\mathcal{H}} = P_i^{\mathcal{C}} + \delta_j \tag{17}$$

The offset $\delta_j$ carries only the semantic difference between a heading and its parent chapter prototype. We treat $P_i^{\mathcal{C}}$ and $\delta_j$ as learnable parameters and define $P_j^{\mathcal{H}}$ only through them in Eq. (17). Accordingly, $P_i^{\mathcal{C}}$ and $\delta_j$ are updated during fine-tuning, whereas $P_j^{\mathcal{H}}$ is reconstructed at every update step. A shift in $P_i^{\mathcal{C}}$ propagates to every heading prototype belonging to the chapter, while a shift in $\delta_j$ repositions only the corresponding heading prototype relative to its parent chapter prototype. Thus, the hierarchy between the two levels is preserved.

### *3.3.2 Prototype-Contrastive Learning*

We perform prototype-contrastive learning at the chapter and heading levels so that the prototypes carry distinct semantic information. To this end, we introduce two triplet losses (Eqs. (18) and (19)). Triplet loss is a metric learning method that learns an embedding space in which an anchor lies closer to a positive sample of the same class than to a negative sample of a different class [40].

$$\mathcal{L}_{HSA}^{\mathcal{C}} = \max(0, \|\bar{h}_x - P_i^{\mathcal{C}+}\|_2 - \|\bar{h}_x - P_{i'}^{\mathcal{C}-}\|_2 + m_1) \tag{18}$$

$$\mathcal{L}_{HSA}^{\mathcal{H}} = \max(0, \|\bar{h}_x - P_j^{\mathcal{H}+}\|_2 - \|\bar{h}_x - P_{j'}^{\mathcal{H}-}\|_2 + m_2) \tag{19}$$

During fine-tuning, the mean-pooled representation $\bar{h}_x$ of the input commodity text serves as the anchor, and the prototypes at each level serve as the positive and negative samples. At the chapter level, the chapter prototype to which $\bar{h}_x$ belongs serves as the positive $P_i^{\mathcal{C}+}$, and the negative $P_{i'}^{\mathcal{C}-}$ is sampled from the remaining chapters $(i' \neq i)$. At the heading level, the prototype of the heading to which $\bar{h}_x$ belongs serves as the positive $P_j^{\mathcal{H}+}$. Because headings subdivide a chapter, the negative $P_{j'}^{\mathcal{H}-}$ is sampled from the headings under the same parent chapter $\pi(j)$. Each triplet loss requires a margin, the minimum difference enforced between the distances from the anchor to the positive and negative prototypes. Because chapters encompass broader semantics than headings, we set the chapter margin $m_1$ larger than the heading margin $m_2$.

Through this mechanism, each commodity text representation is aligned with the prototypes of the chapter and heading to which it belongs. In addition, the prototypes at the same level are semantically separated from one another, as required by the within-level condition.

### *3.3.3 Cross-Level Alignment*

To align the prototypes with the HS code hierarchy, each heading prototype needs to be closest to its parent chapter prototype. Although the triplet losses align the prototypes independently at each level, they cannot guarantee this condition. Therefore, we formulate this condition as a classification task in which each heading prototype identifies its parent chapter. Identifying the parent correctly requires the similarity between the heading prototype and its parent chapter prototype to be the highest among all chapters. Because the prototypes are L2-normalized, these similarities depend only on their directions. Thus, solving this task amounts to aligning the directions of the heading prototype and its parent. Since each prototype is the semantic center of its category, this directional alignment makes the heading prototype semantically closest to its parent chapter prototype.

To implement this task, we define the similarity $s_{i,j}$ between every chapter prototype $P_i^{\mathcal{C}}$ and heading prototype $P_j^{\mathcal{H}}$ as their inner product, which equals cosine similarity (Eq. (20)).

$$s_{i,j} = P_i^{\mathcal{C}} \cdot P_j^{\mathcal{H}} \ \forall i \in \mathcal{C}, j \in \mathcal{H} \tag{20}$$

We then apply a softmax to $s_{i,j}$ over all chapters and define the containment loss $\mathcal{L}_{HSA}^{con}$ as the cross-entropy for this classification task (Eq. (21)). As a result, each heading prototype has the highest similarity to its parent chapter among all chapters, achieving cross-level alignment.

$$\mathcal{L}_{HSA}^{con} = -\frac{1}{|\mathcal{H}|}\sum_{j\in\mathcal{H}} log\frac{exp(s_{\pi(j),j})}{\sum_{i\in\mathcal{C}} exp(s_{i,j})} \tag{21}$$

In summary, $\mathcal{L}_{HSA}^{\mathcal{C}}$ and $\mathcal{L}_{HSA}^{\mathcal{H}}$ in Eqs. (18) and (19) align within-level, whereas $\mathcal{L}_{HSA}^{con}$ in Eq. (21) performs hierarchy-aware alignment between chapter and heading prototypes. These losses are combined as a weighted sum in Eq. (22), where $\lambda \in [0,1]$ is a hyperparameter controlling their relative contributions.

$$\mathcal{L}_{HSA} = \frac{\lambda(\mathcal{L}_{HSA}^{\mathcal{C}} + \mathcal{L}_{HSA}^{\mathcal{H}})}{2} + (1-\lambda)\mathcal{L}_{HSA}^{con} \tag{22}$$

$$\mathcal{L} = \mathcal{L}_{trie} + \mathcal{L}_{HSA} \tag{23}$$

The training objective of TRIE-HSA is defined in Eq. (23). This objective jointly fine-tunes the SLM to generate valid HS codes and to align commodity text representations with the HS code hierarchy.

## 4. Experiments

### *4.1 Experimental Setup*

To evaluate the proposed framework, we use commodity text from import containers collected in the operational environment of the Port of Busan. Commodity text is recorded when a container is unloaded from a vessel and includes information such as the product description, quantity, and packaging unit. Because the HS code of a container is unavailable at the terminal, we generate an HS code for each commodity text with Gemini 2.5 Flash [41]. This process follows Kim et al. [8], who applied Gemini 2.5 to the same type of commodity text and demonstrated the utility of the generated codes in port operations. We regard the generated codes as the ground truth for training and evaluation. The resulting dataset contains 23,355 records and is divided into training, validation, and test sets at a ratio of 60:20:20. The dataset covers 92.7% of HS2 classes, 62.3% of HS4 classes, and 35.9% of HS6 classes (Table 2).

**Table 2. Coverage of HS taxonomy in experimental data**

| Level | Cardinality Taxonomy | Cardinality In Dataset | Coverage |
|---|---|---|---|
| Chapter | 96 | 89 | 92.7% |
| Heading | 1,228 | 765 | 62.3% |
| Subheading | 5,612 | 2,012 | 35.9% |
| **Level** | **Frequency < 10** | **Frequency < 100** | **Frequency 100 +** |
| Chapter | 17 | 28 | 44 |
| Heading | 446 | 274 | 45 |
| Subheading | 1,576 | 409 | 27 |

Whereas previous HSP studies generally constrained evaluation to subsets of HS codes represented in their training data, the present study evaluates all HS codes recorded for the containers in the dataset, thereby establishing a more rigorous setting for assessment of generalization. We adopt accuracy and F1 score as evaluation metrics, as measured at the chapter, heading, and subheading levels, and denote them HS2, HS4, and HS6, respectively.

To ensure a broad and objective evaluation, we select five widely used LLM architectures as backbones, namely Qwen from Alibaba Cloud [42], LLaMA from Meta AI [43], Gemma from Google DeepMind [44], Phi from Microsoft [45], and DeepSeek from DeepSeek AI [46]. These architectures are standard open-source models used extensively in LLM research. Each backbone is fine-tuned with LoRA [47], which freezes the pretrained parameters and updates only the low-rank adapters inserted into its attention layers. We evaluate several scales within each architecture family. Because our objective is to enable high-performance SLM training and inference in resource-constrained environments such as a single-GPU workstation, we limit the evaluation to practically deployable models with at most 14 billion parameters.

We define five experimental configurations to evaluate the proposed framework incrementally. The zero-shot setting (a) comprises (a-1) conventional NTP using only the instruction prompt and (a-2) trie-constrained token prediction without fine-tuning. The fine-tuning setting (b) comprises (b-1) flat prediction of a six-digit HS code using a softmax classification head, (b-2) trie-constrained token prediction without HSA, and (b-3) the complete proposed framework.

These configurations enable a comprehensive comparison of the models' native inference capability, conventional softmax-based classification, and the contributions of the two components of TRIE-HSA. All of the experiments use the same prompt, presented in Appendix 1. The experimental settings and hyperparameters are reported in Appendix 2. Table 3 summarizes the results.

### *4.2 Experimental Results*

Configuration (a-1), which performs HSP using only the native inference capability of an SLM without fine-tuning, yielded generally low accuracy. Most SLMs achieved approximately 10% HS6 accuracy, while even the largest, the Qwen-2.5-14B and Gemma-2-9B models, reached only 31.45 and 21.77%, respectively. Across the 11 evaluated models, the average accuracy was 37.92% at HS2, 19.76% at HS4, and 14.02% at HS6. Nevertheless, performance improved consistently with model scale within the same architecture family. For example, Qwen HS6 accuracy increased from 14.07% with the 3B model to 31.45% with the 14B model, while the evaluated Gemma models improved from 8.69% at 1B to 21.77% at 9B. These findings indicate that the pretrained knowledge of an SLM alone is insufficient for high-performing HSP and that additional fine-tuning is required to learn the relationship between commodity text and the HS taxonomy.

In configuration (a-2), which applies trie-constrained token prediction without fine-tuning, accuracy improved for the Qwen and Gemma families. For example, Gemma-2-9B gained approximately 10% over (a-1). By contrast, LLaMA, Phi, and DeepSeek exhibited substantial performance degradation. This difference arose from architecture-specific tokenization behavior. Qwen and Gemma tokenize continuous digit strings into single-digit tokens, whereas LLaMA and Phi tend to use multi-digit tokens. For example, when instructed to generate "01234567890", Qwen and Gemma output 11 single-digit tokens, while LLaMA-3 and Phi-4-mini tokenize the sequence as four multi-digit tokens, "012", "345", "678", and "90". For the latter architectures, much of the probability mass during HSP was assigned to multi-digit tokens, whereas the trie mask permits only single-digit tokens, which degraded prediction quality. DeepSeek-Distill-7B also generates single-digit tokens but performed poorly because it was trained to produce reasoning before its answer. Its next-token probability mass was concentrated on reasoning tokens, so the trie mask suppressed the reasoning process and left only low-probability digits available for prediction. These findings show that trie-constrained token prediction requires fine-tuning to be effective.

All three fine-tuned configurations, including (b-1) Softmax, (b-2) TRIE, and (b-3) TRIE-HSA, substantially outperformed zero-shot inference, and TRIE-HSA achieved the highest accuracy for every SLM architecture and every hierarchical level. Configuration (b-1), which ignores the HS taxonomy hierarchy, achieved a mean HS6 accuracy of 27.65% and an improvement of 13.63% over (a-1). However, its performance remained limited because it treated each six-digit code as a flat class. Configuration (b-2) achieved a mean HS6 accuracy of 62.65%, improving by 48.63% over (a-1), and consistently outperformed its non-fine-tuned counterpart (a-2). This result indicates that constraining predictions to valid HS codes through the trie mask enables the model to learn the relationship between commodity text and HS codes effectively. Fine-tuning also overcame the architecture-specific limitations observed for LLaMA, Phi, and DeepSeek under (a-2), which structural constraints alone could not resolve.

The complete proposed framework, (b-3) TRIE-HSA, achieved a mean HS6 accuracy of 63.98%, improving by 49.96% over (a-1). Relative to (b-2), which excludes HSA, TRIE-HSA improved the mean HS6 accuracy by 1.33% and the F1 score by 1.94 points. Gemma-2-9B attained the highest HS6 accuracy, at 71.61%, and TRIE-HSA consistently achieved the best results across architectures and hierarchical levels. The contribution of HSA became more pronounced at deeper levels. The gain in accuracy grew from 0.57% at HS2 to 0.98% at HS4 and 1.33% at HS6. The HS6 gain was approximately 2.3 times the HS2 gain. This pattern indicates that HSA is particularly effective for fine-grained prediction, because it explicitly represents the HS taxonomy hierarchy. Zero-shot SLM-based HSP reached 14.02% mean HS6 accuracy, which the proposed framework improved by 49.96%. Further, the LoRA adapters updated

**Table 2. HSP accuracy and F1 score by inference mode and model scale**

| | | Inference Mode | | | | | | | | | |
|---|---|---|---|---|---|---|---|---|---|---|---|
| | | (a) Zero-Shot | | | | (b) Fine-Tuning | | | | | |
| | | (a-1) NTP | | (a-2) TRIE | | (b-1) Softmax | | (b-2) TRIE | | (b-3) TRIE-HSA | |
| Model | | Acc (%) | F1 | Acc | F1 | Acc | F1 | Acc | F1 | Acc | F1 |
| Qwen-2.5-3B | HS2 | 34.43 | 36.72 | **35.67** | **38.01** | 30.59 | 19.71 | 85.91 | **85.63** | **86.30** | 85.40 |
| | HS4 | 15.80 | 19.51 | **16.98** | **20.50** | 24.56 | 14.35 | 73.99 | 72.78 | **74.74** | **73.85** |
| | HS6 | 14.07 | 18.24 | **15.37** | **18.89** | 22.33 | 11.79 | 61.87 | 59.58 | **63.01** | **60.79** |
| Qwen-2.5-7B | HS2 | 64.16 | 64.19 | **64.80** | **64.98** | 38.60 | 29.85 | 87.75 | 87.41 | **88.14** | **87.97** |
| | HS4 | **20.68** | **23.65** | 20.06 | 22.89 | 31.24 | 22.85 | 76.45 | 74.67 | **77.88** | **77.00** |
| | HS6 | 14.58 | **18.54** | **14.99** | 18.40 | 28.13 | 18.46 | 63.90 | 61.62 | **65.28** | **63.98** |
| Qwen-2.5-14B | HS2 | **75.21** | **75.75** | 75.08 | 75.31 | 42.28 | 32.32 | 89.57 | 89.38 | **89.94** | **89.86** |
| | HS4 | 47.74 | 50.76 | **50.03** | **51.73** | 34.96 | 23.62 | 82.00 | 80.96 | **82.17** | **81.79** |
| | HS6 | 31.45 | 35.74 | **34.51** | **36.95** | 30.72 | 17.89 | 69.09 | 66.56 | **69.51** | **67.92** |
| LLaMA-3.2-1B | HS2 | **26.46** | **31.01** | 2.93 | 2.38 | 30.42 | 17.56 | 83.45 | 82.35 | **84.09** | **83.86** |
| | HS4 | **11.92** | **16.96** | 0.62 | 0.43 | 24.62 | 13.04 | 71.38 | 69.91 | **72.47** | **71.40** |
| | HS6 | **10.51** | **14.18** | 0.39 | 0.23 | 22.67 | 11.44 | 60.67 | 58.13 | **61.74** | **59.74** |
| LLaMA-3.2-3B | HS2 | **15.01** | **19.08** | 1.80 | 2.33 | 47.63 | 40.40 | 86.79 | 86.62 | **87.24** | **87.13** |
| | HS4 | **11.18** | **15.14** | 1.09 | 1.62 | 39.14 | 29.58 | 74.95 | 74.35 | **76.17** | **75.67** |
| | HS6 | **9.21** | **12.39** | 1.03 | 1.51 | 34.19 | 23.11 | 62.60 | 60.65 | **64.55** | **62.74** |
| LLaMA-3.1-8B | HS2 | **47.55** | **49.52** | 6.08 | 6.98 | 36.95 | 27.95 | 88.16 | 88.11 | **89.06** | **89.01** |
| | HS4 | **22.01** | **24.58** | 2.21 | 2.82 | 32.52 | 20.45 | 79.30 | 78.87 | **79.85** | **79.45** |
| | HS6 | **11.28** | **13.68** | 1.54 | 1.99 | 29.76 | 16.65 | 66.41 | 64.49 | **67.05** | **65.69** |
| Gemma-3-1B | HS2 | 11.45 | 15.67 | **11.82** | **16.15** | 29.14 | 19.41 | 77.82 | **78.38** | **78.31** | 78.09 |
| | HS4 | 9.46 | **13.96** | **9.63** | 13.39 | 23.42 | 16.64 | 64.27 | 63.91 | **65.92** | **65.90** |
| | HS6 | 8.69 | 12.16 | **9.23** | **12.38** | 21.60 | 14.87 | 54.27 | 52.90 | **55.88** | **54.40** |
| Gemma-2-2B | HS2 | 37.79 | 44.10 | **46.76** | **50.76** | 34.15 | 23.89 | 86.96 | 86.65 | **87.07** | **86.88** |
| | HS4 | 14.99 | **20.38** | **15.20** | 19.92 | 29.05 | 20.59 | 76.60 | 75.16 | **77.35** | **76.89** |
| | HS6 | 13.57 | 17.67 | **14.22** | **18.16** | 25.88 | 16.27 | 63.54 | 60.53 | **64.95** | **63.77** |
| Gemma-2-9B | HS2 | 58.32 | 57.76 | **70.03** | **70.58** | 52.86 | 47.83 | 89.98 | 89.69 | ***90.52** | ***90.47** |
| | HS4 | 41.06 | 42.99 | **50.12** | **52.27** | 42.95 | 34.36 | 83.04 | 81.81 | ***83.41** | ***83.11** |
| | HS6 | 21.77 | 24.86 | **31.32** | **33.85** | 38.69 | 27.49 | 70.11 | 67.90 | ***71.61** | ***69.78** |
| Phi-4-mini-3.8B | HS2 | **34.87** | **39.04** | 1.69 | 1.21 | 38.45 | 28.17 | 83.39 | 83.03 | **84.59** | **84.06** |
| | HS4 | **13.49** | **17.95** | 0.49 | 0.64 | 30.51 | 20.15 | 71.08 | 68.69 | **72.19** | **71.11** |
| | HS6 | **11.58** | **15.30** | 0.34 | 0.40 | 27.36 | 15.99 | 59.32 | 56.15 | **60.86** | **59.05** |
| DeepSeek-Distill-7B | HS2 | **11.82** | **13.69** | 1.58 | 2.26 | 30.76 | 19.21 | 82.04 | 81.78 | **82.79** | **82.53** |
| | HS4 | **9.08** | **13.04** | 1.48 | 2.18 | 25.07 | 14.64 | 68.83 | 67.74 | **70.50** | **69.58** |
| | HS6 | **7.51** | **10.54** | 1.33 | 1.88 | 22.86 | 12.22 | 57.35 | 55.93 | **59.39** | **57.87** |
| Avg | HS2 | **37.92** | **40.59** | 28.93 | 30.09 | 37.44 | 27.85 | 85.62 | 85.37 | 86.19 | **85.93** |
| | HS4 | **19.76** | **23.54** | 15.26 | 17.13 | 30.73 | 20.93 | 74.72 | 73.53 | 75.70 | **75.07** |
| | HS6 | **14.02** | **17.57** | 11.30 | 13.15 | 27.65 | 16.93 | 62.65 | 60.40 | 63.98 | **62.34** |

***: Best overall performance across all models**

less than 0.1% of the backbone parameters and the highest-performing SLM contained fewer than 10 billion parameters. These results demonstrate the practical feasibility of high-performing HSP in resource-constrained environments.

*4.3 Comparison with Benchmarks*

To evaluate the proposed framework against the existing HSP methods, we select seven benchmarks comprising four general-purpose text-embedding models and three HSP-specific architectures. Each benchmark is fine-tuned on the same dataset and under the same experimental conditions as outlined in Appendix 2, following its original study. The text-embedding models are BERT [48], RoBERTa [49], MPNet [50], and FLAN-T5 [51]. BERT is a bidirectional encoder pretrained with masked language modeling and next-sentence prediction, while RoBERTa improves the BERT training strategy using more data and longer training.

Both models also served as benchmarks in the study that proposed SSCNN [7]. MPNet combines masked and permuted language modeling in order to learn token dependencies more effectively. FLAN-T5 is an encoder-decoder model whose instruction tuning improves zero-shot and few-shot generalization across downstream tasks. For HSP, each model embeds commodity text and combines the resulting representation with a classification head.

The three HSP-specific benchmarks are SBERT-MNR, NMT-HL, and SSCNN. SBERT-MNR, proposed by Anggoro et al. [16], fine-tunes a Sentence-BERT encoder with multiple-negatives ranking loss and classifies the resulting representations with a support vector machine (SVM). NMT-HL, proposed by Chen et al. [18], translates commodity text into an HS code with an LSTM encoder-decoder and decodes the chapter, heading, and subheading in three steps under a hierarchical loss. SSCNN, proposed by Zhou et al. [7], combines a shallow-structured CNN for multi-field commodity text with an auxiliary network that incorporates customs-domain knowledge. Because the experimental dataset contains only a single commodity-text field, we omit the auxiliary network and apply only the convolutional neural network (CNN) with the hierarchical focal loss of the original study. For TRIE-HSA, we adopt Gemma-2-9B, which achieved the best performance (see Section 4.2). Table 4 reports the comparison results.

The text-embedding models with classification heads generally performed poorly. BERT and RoBERTa achieved HS6 accuracy of 25.35% and 26.76%, respectively, while MPNet and FLAN-T5 reached only 18.95 and 18.39%. These models treat HSP as flat classification over thousands of subheadings, cannot exploit the label hierarchy, and may not fully capture the meaning of terminology-rich commodity text. Some HSP-specific models achieved competitive performance. Whereas SBERT-MNR obtained only 18.45% HS6 accuracy, NMT-HL and SSCNN achieved 57.05 and 59.67%, respectively. Unlike SBERT-MNR, which learns only representation similarity without modeling the hierarchy directly, NMT-HL and SSCNN decompose prediction by hierarchical level and connect higher-level information to lower-level predictions. This structure enables competitive performance in a highly granular label space. TRIE-HSA outperformed SSCNN, the strongest benchmark, by 13.96, 13.85, and 11.94% in HS2, HS4, and HS6 accuracy, respectively. These findings indicate that an SLM that both constrains generation to valid HS codes and aligns commodity text with the hierarchical semantics of the HS taxonomy offers a more effective approach than existing methods.

*4.4 Interpretability Analysis*

This section analyzes each of the two components of the proposed framework. Both analyses use Gemma-2-9B with TRIE-HSA, the best-performing configuration in Section 4.2.

*4.4.1 Digit-Level Confidence and Prediction Accuracy*

At each digit position, the SLM normalizes its masked logits into a probability distribution. The probability assigned to the predicted digit represents how strongly the SLM concentrates its probability mass on the prediction. Therefore, we define this probability as the confidence of the prediction and examine whether it indicates HS6 accuracy. If the SLM properly captures the meaning of the commodity text, it should assign high confidence to texts that clearly specify the commodity and low confidence to ambiguous ones. Such behavior would indicate that the SLM leverages its language-understanding capability rather than memorized patterns. To test this hypothesis, we predict the HS codes of the test samples and extract the confidence at each digit position. We then group the samples into confidence bins and report the HS6 accuracy of each bin in Fig. 3.

High-confidence bins maintained high HS6 accuracy across all positions. For samples with confidence of at least 99%, HS6 accuracy was 97%, 96%, and 90% at HS2, HS4, and HS6. The corresponding values were 87%, 77%, and 80% in the 95-99% bin and 77%, 74%, and 77% in the 90-95% bin. Every accuracy exceeded the overall HS6 accuracy of 71.61% reported for Gemma-2-9B in Section 4.2, which shows that high-confidence samples substantially outperformed the overall average. Accuracy also decreased monotonically at every position as confidence declined, falling to 36%, 25%, and 37% when confidence was below 50%. This relationship suggests that the SLM derives its confidence from the semantic information in the commodity text. Therefore, the digit-level confidence obtained under trie-constrained token prediction serves as an interpretable indicator of prediction reliability.

**Table 3. Comparison with existing text classification and HSP methods**

| | | HS Level | | | | | |
|---|---|---|---|---|---|---|---|
| | | HS2 | | HS4 | | HS6 | |
| Model | Category | Acc (%) | F1 | Acc (%) | F1 | Acc (%) | F1 |
| BERT | General | 35.15 | 25.21 | 27.90 | 19.49 | 25.35 | 16.24 |
| RoBERTa | General | 34.90 | 25.04 | 29.84 | 20.93 | 26.76 | 17.29 |
| MPNet | General | 23.59 | 11.89 | 20.81 | 10.38 | 18.95 | 9.57 |
| FLAN-T5 | General | 22.39 | 8.72 | 20.08 | 6.95 | 18.39 | 5.88 |
| SBERT-MNR | HSP-Specific | 22.52 | 13.52 | 20.17 | 12.37 | 18.45 | 11.69 |
| NMT-HL | HSP-Specific | 75.40 | 75.05 | 66.43 | 65.10 | 57.05 | 54.91 |
| SSCNN | HSP-Specific | 76.56 | 76.31 | 69.56 | 68.57 | 59.67 | 58.17 |
| TRIE-HSA | Ours | **90.52** | **90.47** | **83.41** | **83.11** | **71.61** | **69.78** |

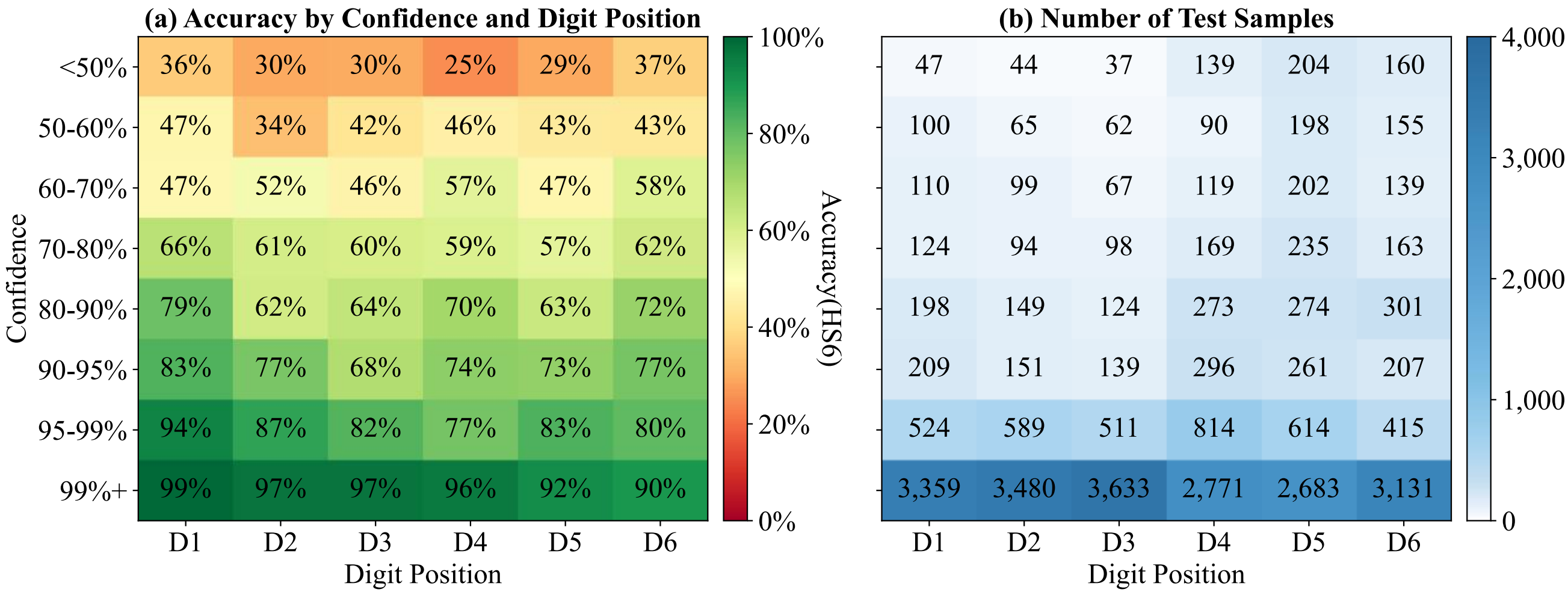

**Fig. 3. Visualization of digit-level confidence**

### *4.4.2 Case Study of Digit-Level Confidence Collapse*

We next examine individual cases to determine whether confidence collapses at the digit position whose deciding information is missing from the commodity text. To this end, we select pairs of a correctly classified case and a misclassified case that share the same actual HS code. Table 5 reports the prediction result and the confidence at each digit position of the predicted HS code. We note that the actual codes were generated from the commodity text by Gemini 2.5 Flash and several details were masked. The generated HS code and the reasoning recorded during generation for each case are presented in Appendix 3.

Case A is a pair whose HS code is 391000, which denotes silicones in primary forms in the HS taxonomy. In Case A.1, the text contains the terms SILICONE PRODUCTS and SILICONE FLUID, and the SLM predicted the correct HS code with confidence of at least 99% at every digit position. In contrast, Case A.2 records only the brand and product name DOWSIL(TM) PMX-1507 without any further description. It is in fact a silicone product of a specific manufacturer, but the SLM failed to capture the meaning of this proprietary term, and its confidence collapsed to 0.5 at the digit position that determines the chapter. This collapse reveals a limitation of the SLM in interpreting brand-specific terms that were not exposed during training.

The HS code of Case B, 020649, denotes other frozen edible offal of swine. Within chapter 02, the heading is determined by the species and by whether the product is meat or edible offal, which separate frozen bovine meat (0202), swine meat (0203), and edible offal (0206). In Case B.1, the text FROZEN PORK STOMACHS states both the species and the part, and the SLM maintained high confidence at every digit position. In contrast, in Case B.2, where only NECKBONES is recorded without the species, confidence collapsed to 0.38 at the digit position that determines the heading.

The HS code of Case C, 401110, denotes new pneumatic tyres for passenger cars. In Case C.1, where the vehicle type is stated as PASSENGER CAR RADIAL, confidence remained consistently high. In Case C.2, which records only PIRELLI BRAND TYRES, confidence stayed above 0.99 up to the heading positions and fell to 0.33 at the subheading

**Table 5. Case studies of digit-level confidence collapse**

| Case | Commodity text | Pred | Actual | Predicted Position $d_1$ | $d_2$ | $d_3$ | $d_4$ | $d_5$ | $d_6$ |
|---|---|---|---|---|---|---|---|---|---|
| A.1 | CONTAINS. 80 DRUMS 16736.000 GROSS WEIGHT KGS NON HAZARDOUS SILICONE PRODUCTS MATERIAL DESCRIPTION PMX-200 SILICONE FLUID 100 CST 200 KGHIGH DENSITY POLYETHYLENE DRUM | 391000 | 391000 | 0.99 | 0.99 | 0.99 | 0.99 | 0.99 | 0.99 |
| A.2 | DOWSIL(TM) PMX-1507 FLUID180 KG DRUM BELGIUM | 382499 | | 0.99 | **0.50** | 0.99 | 0.99 | 0.97 | 0.99 |
| B.1 | FROZEN PORK STOMACHS POUCH BRAND :AVINYO TOTAL QTY :23,000 KG SHIPPER DECLARES ORIGIN SPAIN CIF BUSAN PORT, SOUTH KOREA | 020649 | 020649 | 0.99 | 0.98 | 0.99 | 0.95 | 0.93 | 0.99 |
| B.2 | NECKBONES (FROZ. POLYBLOC) NET WEIGHT :21675.00 KGS GROSS WEIGHT :23864.00 KGS THCD AT DESTINATION PREPAID INLAND HAULAGE HAULAGE AT DESTINATION PREPAID | 020230 | | 0.99 | 0.97 | 0.99 | **0.38** | **0.49** | 0.99 |
| C.1 | 2 X 40 H DC CONTAINER 1584 PIECES MICHELIN BRAND PASSENGER CAR RADIAL LETTER OF CREDIT NO : **** TERMS OF PRICE : FCA EUROPEAN PORT ORIGIN: SPAIN TYRE DETAILS ARE AS PER PURCH | 401110 | 401110 | 0.99 | 0.99 | 0.99 | 0.99 | 0.99 | 0.99 |
| C.2 | PIRELLI BRAND TYRES TEL **** FAX **** CN> **** EMAIL:**** NP> **** EMAIL:**** TAX CODE:690 86 02805 | 401120 | | 0.99 | 0.99 | 0.99 | 0.99 | **0.33** | 0.99 |

red: collapsed confidence position

position that distinguishes tyres by vehicle type. In all three cases, confidence collapsed at the digit position whose deciding information is missing from the commodity text, which provides case-level support for the relationship between confidence and accuracy observed in Section 4.4.1.

#### *4.4.3 Confidence-Based Prediction Truncation*

The case study shows that confidence collapses when the commodity text lacks the information required for HSP. Therefore, the collapse position indicates how far a prediction remains reliable. This subsection quantifies the benefit of truncating the prediction at this position. To this end, we define a confidence threshold $\tau$ and retain the predicted HS code only to the level completed before the first digit position whose confidence falls below $\tau$. For example, if the confidence falls below the threshold at the fifth or sixth digit position, the prediction is retained up to HS4. A truncation is counted as gain when the retained HS levels are correct and the discarded level is incorrect. It is counted as cost when the discarded level is also correct. Samples that are already incorrect at the retained HS levels are excluded from the counts because they fail regardless of truncation. Table 6 reports the gain and cost as $\tau$ varies from 0.9 to 0.3, together with the number and proportion of truncated samples.

**Table 6. Gain and cost of prediction truncation**

| $\tau$ | Truncated | Gain | Cost | Net |
|---|---|---|---|---|
| 0.9 | 2,362 (50.6%) | 869 | 1,413 | -544 |
| 0.8 | 1,859 (39.8%) | 814 | 929 | -115 |
| 0.7 | 1,437 (30.8%) | 707 | 600 | 107 |
| 0.6 | 1,017 (21.8%) | 508 | 360 | 148 |
| 0.5 | 530 (11.3%) | 268 | 141 | 127 |
| 0.4 | 244 (5.2%) | 130 | 46 | 84 |
| 0.3 | 49 (1.0%) | 20 | 8 | 12 |

The highest threshold $\tau = 0.90$ truncated a prediction whenever any digit position fell below 0.90. It truncated 50.6% of the test set with a gain of 869 and a cost of 1,413, so the cost exceeded the gain by 544. Such a high threshold discarded correct predictions more often than incorrect ones. By contrast, $\tau = 0.40$ truncated only 244 samples. Although the gain of 130 far exceeded the cost of 46, the net gain reached only 84. $\tau = 0.60$ balanced these two effects. It truncated 1,017 samples or 21.8% of the test set with a gain of 508 and a cost of 360, and the resulting net gain of 148, which corresponds to 3.2% of the test set, was the largest in Table 6. These results show that confidence-based truncation indicates how far a prediction can be trusted for the noisy commodity text collected at container terminals. We discuss the practical implications of this result in Section 5.1.

#### *4.4.4 Semantic Separation in the Representation Space*

We next examine whether HSA establishes the HS code hierarchy in the representation space. To this end, we visualize the prototypes and the commodity text representations in that space during fine-tuning. To keep the figure legible, we plot the 14 most frequent chapters and their headings.

Fig. 4 shows the chapter prototypes, heading prototypes, and representations of validation commodity texts not used for fine-tuning. The representations are projected into two dimensions using t-distributed stochastic neighbor embedding (t-SNE), with each panel showing the state after one completed epoch. Before fine-tuning, the prototypes are unseparated and the commodity text representations are not yet aligned with them. As fine-tuning proceeded, the chapter prototypes became semantically separated by discernible distances, while each heading prototype and validation sample formed a semantic cluster around its parent chapter prototype. This pattern indicates that the two HSA triplet losses and the containment loss jointly learn both within-level category discrimination and cross-level containment relationships. Because this organization separates semantic information more clearly as the hierarchy deepens, it supports the finding in Section 4.2 that HSA improved accuracy more at the deeper levels. Furthermore, Fig. 4 shows only samples excluded from fine-tuning, which indicates that the hierarchy-aware alignment observed here extends to unseen commodity text.

These analyses show that the components of TRIE-HSA act on different aspects of the task. Trie-constrained token prediction allows the SLM to assign confidence according to the semantic information in commodity text, while HSA

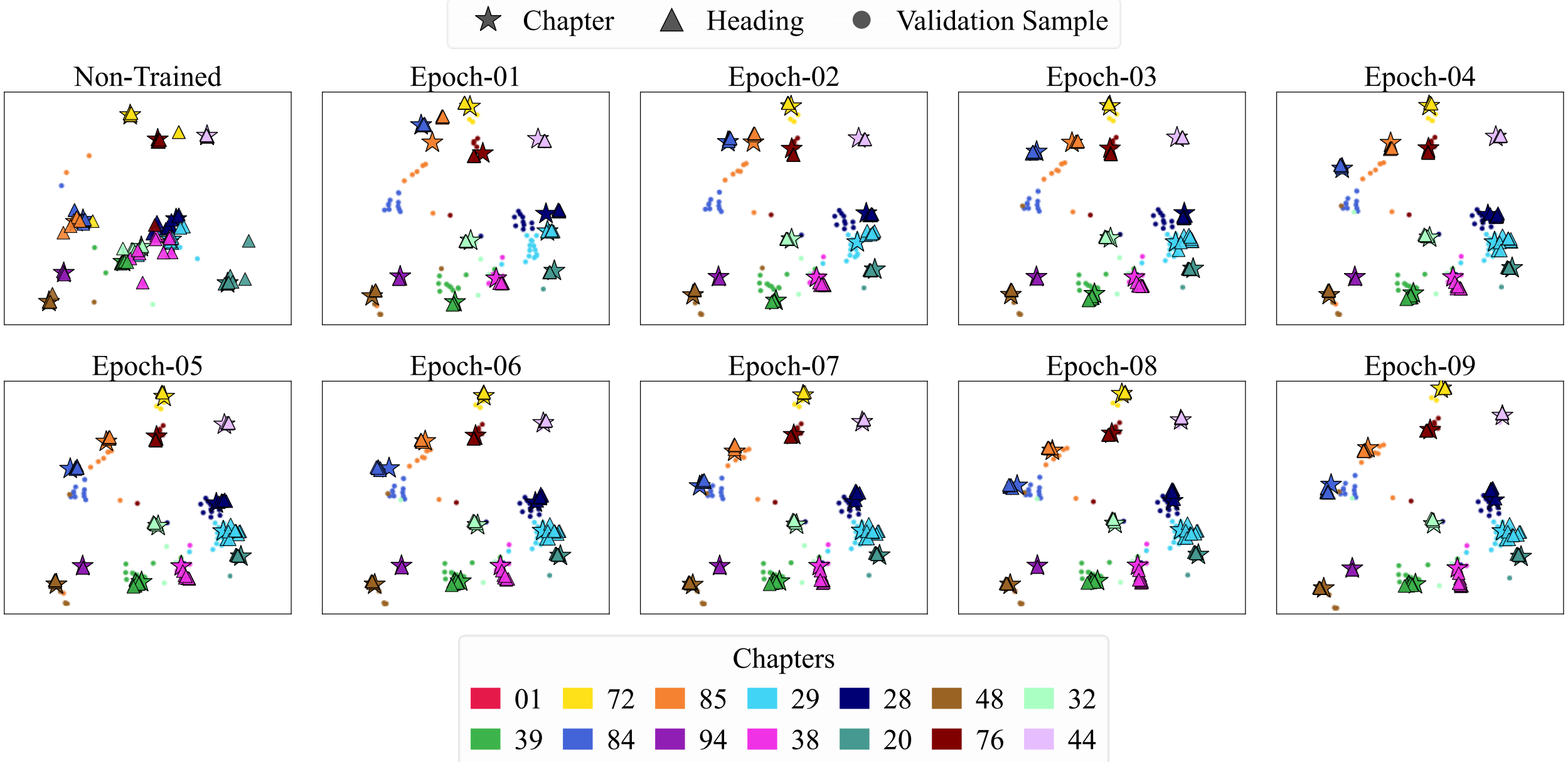

**Fig. 4. Hierarchy-aware alignment across fine-tuning epochs**

arranges the prototypes and the commodity text representations along the HS code hierarchy.

*4.5 Computational Efficiency*

This section examines the computational cost of the proposed framework. The best-performing backbone, Gemma-2-9B, contains fewer than 10 billion parameters, and its LoRA adapters update less than 0.1% of them. To verify that this scale is practical, we measure the fine-tuning and prediction cost on a single consumer-grade GPU with 32 GB of memory, an NVIDIA GeForce RTX 5090. We fine-tune the backbone under the experimental settings reported in Appendix 2. To measure the prediction time, we sample 100 commodity texts at random from the experimental data and repeat the procedure 10 times with a batch size of one. We also run conventional NTP on the same backbone for comparison, capping its generation at six tokens so that both configurations produce an output of the same length. NTP uses the pretrained weights without fine-tuning. Table 7 reports the results.

**Table 7. Computational efficiency**

| Avg. Tokens=133.60 | **TRIE-HSA** | **NTP** |
|---|---|---|
| Epochs | 20 | - |
| Params | 9.24B | |
| Trainable Params | 0.0092B | - |
| Training Time per Epoch | 17.32 min | - |
| Total Training Time | 155.88 min | - |
| Prediction Time-Min | 0.311 sec | 0.343 sec |
| Prediction Time-Max | 0.494 sec | 0.481 sec |
| Prediction Time-Avg | 0.335 sec | 0.358 sec |

Fine-tuning Gemma-2-9B with TRIE-HSA took 17.32 minutes per epoch, finishing in 155.88 minutes over nine epochs with early stopping. At prediction time, the model required 0.335 seconds per commodity text on average, and the difference from NTP remained within 0.023 seconds, which indicates that the proposed framework added no measurable cost to prediction. These measurements confirmed that both fine-tuning and prediction run on a single-GPU workstation.

## 5. Discussion

*5.1 Practical Implications*

The first implication of this study concerns the computational resources that HSP requires. As reported in Section 4.5, both fine-tuning and prediction ran on a single consumer-grade GPU, and the trie-constraint added no measurable cost at prediction time. Therefore, the proposed framework alleviates the high cost of LLM-based HSP noted in previous studies [2,8] and allows SLM-based HSP to operate in resource-constrained environments. Operating on a single GPU also permits on-premise deployment. This capability not only extends the proposed framework to domains that handle commercially sensitive text, but also allows each terminal to fine-tune the SLM on locally collected commodity text. Because commodity composition differs across ports, this local fine-tuning specializes the SLM to the commodities handled at each terminal. In this way, an SLM alone can convert unstructured commodity text into the structured codes that previous studies used to improve terminal operations [8,23].

The second implication is the applicability of the proposed framework to other domains. Studies that predict text under a hierarchical taxonomy have been conducted in various fields. Coutinho et al. [29] generated ICD-10 codes from death-certificate text for clinical coding, and Singh and Diao [52] categorized product descriptions under the UNSPSC taxonomy for procurement. Safikhani et al. [53] adopted pretrained language models to assign hierarchical occupation codes to free-text job descriptions. These tasks share the structure of HSP, in which unstructured text needs to be mapped onto a code whose levels carry hierarchical meaning. Therefore, the proposed framework can be applied to such tasks beyond HSP.

The third implication is that the proposed framework enables selective use of the predictions. In operational environments such as container terminals, commodity text is often incomplete and noisy. Although the proposed framework guarantees that every prediction is a valid HS code, a prediction can still be incorrect and lead to erroneous operational decisions. Therefore, practical deployment requires a criterion for using each prediction selectively. The digit-level confidence provides such a criterion. As reported in Section 4.4, the confidence was closely associated with prediction accuracy, and truncation below an appropriate threshold proved effective. These characteristics allow a deployed framework to determine the level to which each prediction is adopted and to refer the rest to expert review. This selective use enables effective decision making in daily operations.

*5.2 Limitations and Future Research*

The commodity text used in this study was collected from an operational container terminal at the Port of Busan. Because the HS code of commodity text is unavailable at the container terminal, we followed previous studies and regarded the HS codes that an LLM assigned to the commodity text as the ground truth [8]. Therefore, the experimental results in Section 4.2 can be interpreted as measuring how similarly the SLM and the LLM predict HS codes. Although this evaluation is consistent with the purpose of this study, namely determining whether an SLM can substitute for an LLM, practical deployment of the proposed framework requires extensive validation on customs-classified commodity text.

Although TRIE-HSA reduces the computational burden of HSP, it still depends on fine-tuning with labeled HS codes. Because container terminals generally do not manage these codes, the labels have to be prepared separately. In this study, the labels were generated with an LLM in an offline step, and this labeling incurred a cost. However, this step is performed once before deployment and requires minimal computing infrastructure, whereas the SLM runs continuously in the operational environment. Thus, the overall cost of HSP remains low. Where cooperation with customs authorities is possible, customs-classified records of the corresponding country can be utilized for fine-tuning. Releasing SLMs fine-tuned with the framework as open weights is also a promising direction for future research, since it would allow terminals to adopt the framework without labels of their own and further broaden its applicability.

## 6. Conclusion

This study proposed TRIE-HSA for HSP in resource-constrained port-logistics environments. The framework addresses two major weaknesses of SLMs, specifically by eliminating invalid-code hallucinations and aligning commodity text with the HS code hierarchy. In experiments using commodity text collected from a container terminal, an SLM fine-tuned with TRIE-HSA improved HS6 accuracy by 49.96% over zero-shot inference and outperformed the existing benchmarks by 11.94%. These results arose from three aspects of the framework. First, the trie-constrained token prediction guarantees that every prediction is valid under the HS taxonomy and thereby improves HSP accuracy. Second, HSA aligns commodity text with the HS code hierarchy, meaning that explicitly representing hierarchical semantics can contribute to prediction performance. Third, TRIE-HSA achieves these results with a 9B-parameter SLM, offering potential practical applicability in resource-constrained environments without reliance on large-scale LLMs.

## Appendix 1. Prompt Design

**Table A1. Prompt used for HS code prediction**

| Prompt |
|---|
| You are an HS code classification expert. Classify the given commodity text into a six-digit HS code. Output ONLY the code in 6-digit format, nothing else.<br>commodity text: $\{x\}$ |

**Appendix 2. Experimental Settings and Hyperparameters**

For the four general-purpose text-embedding models, BERT, RoBERTa, MPNet, and FLAN-T5, we used the pretrained weights released on Hugging Face as the encoder and attached a classification head, which was the fine-tuning target. FLAN-T5 is an encoder-decoder model, and we used its encoder alone. SBERT-MNR and NMT-HL were reimplemented following the descriptions in their original studies, and SSCNN was trained with the source code released by its authors.

**Table A2. Experimental settings and hyperparameter search space**

| Name | Value | Name | Value |
|---|---|---|---|
| Observations | 23,355 | Epochs | 20 |
| Split Ratio | Train: 60%/Validation: 20%/Test: 20% | Early Stopping | 3 |
| Optimizer | AdamW | Batch Size | 32 |

| Model | Category | Type | Training | Hyperparameters | Range | Step | Select |
|---|---|---|---|---|---|---|---|
| BERT | General | Transformer Encoder | FT | learning rate | [0.0001, 0.001] | 0.0005 | 0.0001 |
| | | | | dropout | [0.1, 0.3] | 0.1 | 0.1 |
| | | | | weight decay | [0, 0.01] | 0.01 | 0.01 |
| RoBERTa | General | Transformer Encoder | FT | learning rate | [0.0001, 0.001] | 0.0005 | 0.0001 |
| | | | | dropout | [0.1, 0.3] | 0.1 | 0.1 |
| | | | | weight decay | [0, 0.01] | 0.01 | 0.01 |
| MPNet | General | Transformer Encoder | FT | learning rate | [0.0001, 0.001] | 0.0005 | 0.0001 |
| | | | | dropout | [0.1, 0.3] | 0.1 | 0.1 |
| | | | | weight decay | [0, 0.01] | 0.01 | 0.01 |
| FLAN-T5 | General | Transformer Encoder-Decoder | FT | learning rate | [0.0001, 0.001] | 0.0005 | 0.001 |
| | | | | dropout | [0.1, 0.3] | 0.1 | 0.1 |
| | | | | weight decay | [0, 0.01] | 0.01 | 0.01 |
| SBERT-MNR | HSP-Specific | Transformer Encoder | FT | backbone | SBERT | - | SBERT |
| | | | | kernel (SVM) | RBF | - | RBF |
| | | | | C (SVM) | 1.0 | - | 1.0 |
| | | | | max iter (SVM) | [100, 1,000] | 100 | 1,000 |
| | | | | learning rate | [0.0001, 0.001] | 0.0005 | 0.005 |
| | | | | dropout | [0.1, 0.3] | 0.1 | 0.3 |
| NMT-HL | HSP-Specific | RNN+Attention | Full Training | hidden size | [128, 512] | 128 | 256 |
| | | | | $HL\ \alpha$ | [0.5, 1.5] | 0.5 | 1 |
| | | | | $HL\ \beta$ | [0.3, 0.5] | 0.1 | 0.5 |
| | | | | learning rate | [0.0001, 0.001] | 0.0005 | 0.0005 |
| | | | | dropout | [0.1, 0.3] | 0.1 | 0.1 |
| SSCNN | HSP-Specific | CNN | Full Training | hidden size | [200, 300] | 100 | 300 |
| | | | | filters | [150, 250] | 50 | 250 |
| | | | | $HL\ \beta$ | [0.3, 0.5] | 0.1 | 0.5 |
| | | | | $Focal\ \alpha$ | [0.25, 0.5] | 0.25 | 0.5 |
| | | | | $Focal\ \gamma$ | [0.5, 2.0] | 0.5 | 0.5 |
| | | | | learning rate | [0.0001, 0.001] | 0.0005 | 0.001 |
| | | | | dropout | [0.1, 0.3] | 0.1 | 0.3 |
| TRIE-HSA | Ours | SLM | PEFT | r (LoRA) | [8, 16] | 8 | 8 |
| | | | | alpha (LoRA) | [16, 32] | 16 | 16 |
| | | | | LoRA dropout | [0.01, 0.05] | 0.01 | 0.01 |
| | | | | $\lambda$ | [0.3, 0.6] | 0.1 | 0.5 |
| | | | | $m_1$ | [1, 2] | 1 | 1 |
| | | | | $m_2$ | [0.5, 1] | 0.5 | 0.5 |
| | | | | learning rate | [0.0001, 0.001] | 0.0005 | 0.001 |
| | | | | dropout | [0.1, 0.3] | 0.1 | 0.1 |

FT: Fine Tuning
PEFT: Parameter-Efficient Fine-Tuning

### Appendix 3. Ground-Truth Generation for the Case-Study Samples

Table A3 lists the generated codes and the reasons that Gemini 2.5 Flash recorded during the ground-truth generation described in Section 4.1. For most samples, Gemini 2.5 Flash identified the commodity from evidence stated in the text, such as the material term “SILICONE FLUID” in Case A, the species and part in “FROZEN PORK STOMACHS” in Case B, and the vehicle type in “PASSENGER CAR RADIAL” in Case C. In contrast, for the collapsed samples of Cases B and C, the recorded reasons involve presumption. The species of “NECKBONES” is recorded as presumed swine, and the vehicle type of “PIRELLI BRAND TYRES” is inferred from the brand rather than stated in the text.

**Table A3. Reasons recorded during ground-truth generation based on Gemini 2.5-Flash**

| Case | Commodity text | Recorded Reason | Generated HS code |
|---|---|---|---|
| A.1 | CONTAINS. 80 DRUMS 16736.000 GROSS WEIGHT KGS NON HAZARDOUS SILICONE PRODUCTS MATERIAL DESCRIPTION PMX-200 SILICONE FLUID 100 CST 200 KGHIGH DENSITY POLYETHYLENE DRUM | The cargo is explicitly described as “SILICONE FLUID” and “SILICONE PRODUCTS”. HS 3910 covers silicones in primary forms, which includes silicone fluids. The product name “PMX-200” further confirms a silicone product. | 391000 |
| A.2 | DOWSIL(TM) PMX-1507 FLUID180 KG DRUM BELGIUM | The cargo is identified as DOWSIL(TM) PMX-1507 FLUID, which is a silicone blend (dimethicone/dimethiconol). Silicones in primary forms, including liquid forms, are classified under heading 3910. | |
| B.1 | FROZEN PORK STOMACHS POUCH BRAND :AVINYO TOTAL QTY :23,000 KG SHIPPER DECLARES ORIGIN SPAIN CIF BUSAN PORT, SOUTH KOREA | The cargo is explicitly stated as “FROZEN PORK STOMACHS”. HS 0206 covers edible offal of bovine animals, swine, and other species; 0206.49, other frozen swine offal, exactly matches frozen pork stomachs. | 020649 |
| B.2 | NECKBONES (FROZ. POLYBLOC) NET WEIGHT :21675.00 KGS GROSS WEIGHT :23864.00 KGS THCD AT DESTINATION PREPAID INLAND HAULAGE HAULAGE AT DESTINATION PREPAID | “NECKBONES” and “FROZ.” indicate edible offal of a frozen swine (*presumed*). Heading 0206 includes frozen edible offal, and 0206.49, other frozen swine offal, is the most appropriate subheading. | |
| C.1 | 2 X 40 H DC CONTAINER 1584 PIECES MICHELIN BRAND PASSENGER CAR RADIAL LETTER OF CREDIT NO : **** TERMS OF PRICE : FCA EUROPEAN PORT ORIGIN: SPAIN TYRE DETAILS ARE AS PER PURCH | The commodity is clearly described by “TIRE” and “PASSENGER CAR RADIAL”, and indicates new pneumatic tyres for passenger cars, HS 4011.10. | 401110 |
| C.2 | PIRELLI BRAND TYRES TEL **** FAX **** CN> **** EMAIL:**** NP> **** EMAIL:**** TAX CODE:690 86 02805 | The text states “PIRELLI BRAND TYRES”, indicating new pneumatic rubber tyres under heading 4011. Since PIRELLI mainly produces tyres for passenger cars, classification under 4011.10 is appropriate. | |

## References

[1] H. van der Heijden, Predicting industry sectors from financial statements: An illustration of machine learning in accounting research, Br. Account. Rev. 54 (5) (2022) 101096, https://dx.doi.org/10.1016/j.bar.2022.101096.

[2] P. Yuvraj, S. Devarakonda, ATLAS: Benchmarking and adapting LLMs for global trade via harmonized tariff code classification, arXiv preprint arXiv:2509.18400 (2025), https://dx.doi.org/10.48550/arXiv.2509.18400.

[3] S. Nath, S. Wadhwa, L. Perez, Domain-adaptive small language models for structured tax code prediction, arXiv preprint arXiv:2507.10880 (2025), https://dx.doi.org/10.48550/arXiv.2507.10880.

[4] L. Qi, Q. Zhang, X. Lin, J. Zhang, M. Liao, Attribute knowledge and KBGAT for predicting the accuracy of the harmonized system code for classifying import and export commodities, Sci. Rep. 15 (1) (2025) 43504, https://dx.doi.org/10.1038/s41598-025-16580-7.

[5] Z. Navasardyan, Interpretable and generalizable HTS code classification framework, Econ. Finance Account. 1 (13) (2024) 140, https://dx.doi.org/10.59503/29538009-2024.1.13-140.

[6] F. Altaheri, K. Shaalan, Exploring machine learning models to predict harmonized system code, in: Information Systems, EMCIS 2019, Springer, Cham, 2020, pp. 291–303, https://doi.org/10.1007/978-3-030-44322-1_22.

[7] C. Zhou, C. Che, X.S. Zhang, Q. Zhang, D. Zhou, Harmonized system code prediction of import and export commodities based on hybrid convolutional neural network with auxiliary network, Knowl.-Based Syst. 256 (2022) 109836, https://dx.doi.org/10.1016/j.knosys.2022.109836.

[8] M. Kim, T. Kim, T. Park, H. Park, H. Bae, Generative AI and machine learning collaboration for container dwell time prediction via data standardization, Transp. Res. E Logist. Transp. Rev. 216 (2026) 105171, https://doi.org/10.1016/j.tre.2026.105171.

[9] T. Lan, Y. Yang, Q. Jia, L. Zhu, H. Jiang, H. Zhu, et al., HSCodeComp: A realistic and expert-level agent benchmark for hierarchical rule application, in: Proc. 64th Annu. Meet. Assoc. Comput. Linguist. (2026), https://dx.doi.org/10.18653/v1/2026.acl-long.937.

[10] L. Ding, Z. Fan, D. Chen, Auto-categorization of HS code using background net approach, Procedia Comput. Sci. 60 (2015) 1462–1471, https://dx.doi.org/10.1016/j.procs.2015.08.224.

[11] M. Spichakova, H.-M. Haav, Application of machine learning for assessment of HS code correctness, Balt. J. Mod. Comput. 8 (4) (2020) 698–718, https://dx.doi.org/10.22364/bjmc.2020.8.4.13.

[12] E. Lee, S. Kim, S. Kim, S. Park, M. Cha, S. Jung, et al., Classification of goods using text descriptions with sentences retrieval, arXiv preprint arXiv:2111.01663 (2021), https://dx.doi.org/10.48550/arXiv.2111.01663.

[13] O. Amel, S. Stassin, S.A. Mahmoudi, X. Siebert, Multimodal approach for harmonized system code prediction, in: 31st Eur. Symp. Artif. Neural Netw. Comput. Intell. Mach. Learn., (2023), pp. 181–186, https://dx.doi.org/10.14428/esann/2023.ES2023-165.

[14] M. He, X. Wang, C. Zou, B. Dai, L. Jin, A commodity classification framework based on machine learning for analysis of trade declaration, Symmetry 13 (6) (2021) 964, https://dx.doi.org/10.3390/sym13060964.

[15] H. Sun, C. Zhou, C. Che, Customs commodity classification method based on the fusion of text sequence and graph information, Expert Syst. 42 (6) (2025) e70057, https://dx.doi.org/10.1111/exsy.70057.

[16] A.W. Anggoro, P. Corcoran, D. De Widt, Y. Li, Harmonized system code classification using supervised contrastive learning with sentence BERT and multiple negative ranking loss, Data Technol. Appl. 59 (2) (2025) 276–301, https://dx.doi.org/10.1108/DTA-01-2024-0052.

[17] N.T. Binh, H.A. Nguyen, P.N. Linh, N.L. Giang, T.N. Thang, Attentive RNN for HS code hierarchy classification on Vietnamese goods declaration, in: D.-T. Tran, G. Jeon, T.D.L. Nguyen, J. Lu, T.-D. Xuan (Eds.), Intelligent Systems and Networks, Lecture Notes in Networks and Systems, vol. 243, Springer, Singapore, 2021, pp. 298–304, https://dx.doi.org/10.1007/978-981-16-2094-2_37.

[18] X. Chen, S. Bromuri, M. van Eekelen, Neural machine translation for harmonized system codes prediction, in: 2021 6th Int. Conf. Mach. Learn. Technol., ICMLT, 2021, pp. 158–163, https://dx.doi.org/10.1145/3468891.3468915.

[19] Shubham, A. Arya, S. Roy, S. Jonnala, An ensemble-based approach for assigning text to correct harmonized system code, in: 2023 Int. Conf. Artif. Intell. Smart Commun., AISC, 2023, pp. 35–41, https://dx.doi.org/10.1109/AISC56616.2023.10085512.

[20] S.V. Kandappareddigari, S. Jagadish, G. Verma, I. Contreras, C. Dignam, A. Srivastava, et al., Operationalization of machine learning with serverless architecture: An industrial implementation for harmonized system code prediction, arXiv preprint arXiv:2602.17102 (2026), https://dx.doi.org/10.48550/arXiv.2602.17102.

[21] E. Lee, S. Kim, S. Kim, S. Jung, H. Kim, M. Cha, Explainable product classification for customs, ACM Trans. Intell. Syst. Technol. 15 (2) (2024) 1–24, https://dx.doi.org/10.1145/3635158.

[22] I. Marra De Artiñano, F. Riottini Depetris, C. Volpe Martincus, Automatic product classification in international trade: Machine learning and large language models, Rev. Int. Econ. 34 (1) (2026) 3–19, http://dx.doi.org/10.1111/roie.70009.

[23] Y. Xie, D.-P. Song, J. Dong, Y. Feng, Predicting out-terminals for imported containers at seaports using machine learning: Incorporating unstructured data and measuring operational costs due to misclassifications, Transp. Res. E Logist. Transp. Rev. 202 (2025) 104331, https://dx.doi.org/10.1016/j.tre.2025.104331.

[24] C. Hokamp, Q. Liu, Lexically constrained decoding for sequence generation using grid beam search, in: Proc. 55th Annu. Meet. Assoc. Comput. Linguist. 2017, pp. 1535–1546, https://dx.doi.org/10.18653/v1/P17-1141.

[25] S. Geng, M. Josifoski, M. Peyrard, R. West, Grammar-constrained decoding for structured NLP tasks without finetuning, in: Proc. 2023 Conf. Empir. Methods Nat. Lang. Process., 2023, pp. 10932–10952, https://dx.doi.org/10.18653/v1/2023.emnlp-main.674.

[26] X. Lu, P. West, R. Zellers, R. Le Bras, C. Bhagavatula, Y. Choi, NeuroLogic decoding: (Un)supervised neural text generation with predicate logic constraints, in: Proc. 2021 Conf. North Am. Chapter Assoc. Comput.

Linguist.: Hum. Lang. Technol., 2021, pp. 4288–4299, https://dx.doi.org/10.18653/v1/2021.naacl-main.339.

[27] T. Scholak, N. Schucher, D. Bahdanau, PICARD: Parsing incrementally for constrained auto-regressive decoding from language models, in: Proc. 2021 Conf. Empir. Methods Nat. Lang. Process., 2021, pp. 9895–9901, https://dx.doi.org/10.18653/v1/2021.emnlp-main.779.

[28] B.T. Willard, R. Louf, Efficient guided generation for large language models, arXiv preprint arXiv:2307.09702 (2023), https://dx.doi.org/10.48550/arXiv.2307.09702.

[29] I. Coutinho, G.M. Correia, B. Martins, A. Moreira, A. Peralta-Santos, ICD coding of death certificates with generative language models, PLOS Digit. Health 5 (2) (2026) e0001245, https://dx.doi.org/10.1371/journal.pdig.0001245.

[30] N. De Cao, G. Izacard, S. Riedel, F. Petroni, Autoregressive entity retrieval, in: Int. Conf. Learn. Represent., ICLR, 2021, https://openreview.net/forum?id=5k8F6UU39V.

[31] Z. Su, I. Katsman, Y. Wang, R. He, L. Heldt, R. Keshavan, et al., Vectorizing the trie: Efficient constrained decoding for LLM-based generative retrieval on accelerators, in: Proc. 32nd ACM SIGKDD Conf. Knowl. Discov. Data Min., KDD, 2026, http://dx.doi.org/10.1145/3770855.3818506.

[32] X. Song, L. Huang, H. Xue, S. Hu, Supervised prototypical contrastive learning for emotion recognition in conversation, in: Proc. 2022 Conf. Empir. Methods Nat. Lang. Process., 2022, pp. 5197–5206, https://dx.doi.org/10.18653/v1/2022.emnlp-main.347.

[33] G. Cui, S. Hu, N. Ding, L. Huang, Z. Liu, Prototypical verbalizer for prompt-based few-shot tuning, in: Proc. 60th Annu. Meet. Assoc. Comput. Linguist. 2022, pp. 7014–7024, https://dx.doi.org/10.18653/v1/2022.acl-long.483.

[34] G. Wang, Y. Du, Y. Jiang, LSPCL: Label-specific supervised prototype contrastive learning for multi-label text classification, Knowl.-Based Syst. 309 (2025) 112887,https://dx.doi.org/10.1016/j.knosys.2024.112887.

[35] Z. Wang, P. Wang, L. Huang, X. Sun, H. Wang, Incorporating hierarchy into text encoder: A contrastive learning approach for hierarchical text classification, in: Proc. 60th Annu. Meet. Assoc. Comput. Linguist. 2022, pp. 7109–7119, https://dx.doi.org/10.18653/v1/2022.acl-long.491.

[36] S.C.L. Yu, J. He, V. Gutiérrez-Basulto, J.Z. Pan, Instances and labels: Hierarchy-aware joint supervised contrastive learning for hierarchical multi-label text classification, in: Findings Assoc. Comput. Linguist.: EMNLP 2023, 2023, pp. 8858–8875, https://dx.doi.org/10.18653/v1/2023.findings-emnlp.594.

[37] S. Kim, B. Jeong, S. Kwak, HIER: Metric learning beyond class labels via hierarchical regularization, in: 2023 IEEE/CVF Conf. Comput. Vis. Pattern Recognit., CVPR, 2023, pp. 19903–19912, https://dx.doi.org/10.1109/CVPR52729.2023.01906.

[38] J. Zhang, Y. Li, F. Shen, C. Xia, H. Tan, Y. He, Hierarchy-aware and label balanced model for hierarchical text classification, Knowl.-Based Syst. 300 (2024) 112153, https://dx.doi.org/10.1016/j.knosys.2024.112153.

[39] T.B. Brown, B. Mann, N. Ryder, M. Subbiah, J. Kaplan, P. Dhariwal, et al., Language models are few-shot learners, in: Adv. Neural Inf. Process. Syst. 33, 2020, pp. 1877–1901.

[40] F. Schroff, D. Kalenichenko, J. Philbin, FaceNet: A unified embedding for face recognition and clustering, in: 2015 IEEE Conf. Comput. Vis. Pattern Recognit., CVPR, 2015, pp. 815–823, https://dx.doi.org/10.1109/CVPR.2015.7298682.

[41] G. Comanici, E. Bieber, M. Schaekermann, I. Pasupat, N. Sachdeva, I. Dhillon, et al., Gemini 2.5: Pushing the frontier with advanced reasoning, multimodality, long context, and next generation agentic capabilities, arXiv preprint arXiv:2507.06261 (2025), https://dx.doi.org/10.48550/arXiv.2507.06261.

[42] Qwen Team, A. Yang, B. Yang, B. Zhang, B. Hui, B. Zheng, et al., Qwen2.5 technical report, arXiv preprint arXiv:2412.15115 (2025), https://dx.doi.org/10.48550/arXiv.2412.15115.

[43] A. Grattafiori, A. Dubey, A. Jauhri, A. Pandey, A. Kadian, A. Al-Dahle, et al., The Llama 3 herd of models, arXiv preprint arXiv:2407.21783 (2024), https://dx.doi.org/10.48550/arXiv.2407.21783.

[44] Gemma Team, A. Kamath, J. Ferret, S. Pathak, N. Vieillard, R. Merhej, et al., Gemma 3 technical report, arXiv preprint arXiv:2503.19786 (2025), https://dx.doi.org/10.48550/arXiv.2503.19786.

[45] Microsoft, A. Abouelenin, A. Ashfaq, A. Atkinson, H. Awadalla, N. Bach, et al., Phi-4-Mini technical report: Compact yet powerful multimodal language models via mixture-of-LoRAs, arXiv preprint arXiv:2503.01743 (2025), https://dx.doi.org/10.48550/arXiv.2503.01743.

[46] D. Guo, D. Yang, H. Zhang, J. Song, P. Wang, Q. Zhu, et al., DeepSeek-R1 incentivizes reasoning in LLMs through reinforcement learning, Nature 645 (8081) (2025) 633–638, https://dx.doi.org/10.1038/s41586-025-09422-z.

[47] E.J. Hu, Y. Shen, P. Wallis, Z. Allen-Zhu, Y. Li, S. Wang, et al., LoRA: Low-rank adaptation of large language models, in: Int. Conf. Learn. Represent., ICLR, 2022, https://openreview.net/forum?id=nZeVKeeFYf9.

[48] J. Devlin, M.-W. Chang, K. Lee, K. Toutanova, BERT: Pre-training of deep bidirectional transformers for language understanding, in: Proc. 2019 Conf. North Am. Chapter Assoc. Comput. Linguist.: Hum. Lang. Technol., 2019, pp. 4171–4186, https://dx.doi.org/10.18653/v1/N19-1423.

[49] Y. Liu, M. Ott, N. Goyal, J. Du, M. Joshi, D. Chen, et al., RoBERTa: A robustly optimized BERT pretraining approach, arXiv preprint arXiv:1907.11692 (2019), https://dx.doi.org/10.48550/arXiv.1907.11692.

[50] K. Song, X. Tan, T. Qin, J. Lu, T.-Y. Liu, MPNet: Masked and permuted pre-training for language understanding, in: Adv. Neural Inf. Process. Syst. 33, 2020, pp. 16857–16867.

[51] H.W. Chung, L. Hou, S. Longpre, B. Zoph, Y. Tay, W. Fedus, et al., Scaling instruction-finetuned language models, J. Mach. Learn. Res. 25 (70) (2024) 1–53.

[52] A. Singh, Y. Diao, Leveraging large language models for optimized item categorization using UNSPSC taxonomy, Int. J. Cybern. Inform. 13 (6) (2024) 1–10, https://dx.doi.org/10.5121/ijci.2024.130601.